\documentclass[aps,prd,twocolumn,superscriptaddress]{revtex4-1}

\usepackage[separate-uncertainty=true]{siunitx}
\DeclareSIUnit{\sqrthz}{\ensuremath{\sqrt{\textrm{\hertz}}}} 

\usepackage[tight,footnotesize]{subfigure}
\usepackage{graphicx}
\usepackage{amsmath}

\newcommand{\ssm}{Sagnac speed meter}
\newcommand{\mi}{$\textrm{M}_{\textrm{I}}$}
\newcommand{\rt}{$\textrm{R}_{\textrm{T}}$}
\newcommand{\lminus}{$L_{\left(-\right)}$}
\newcommand{\fp}{Fabry-P\'{e}rot}

\begin{document}

\title{Control of a velocity-sensitive audio-band quantum non-demolition interferometer}

\author{S. S. Leavey}
  \email{s.leavey.1@research.gla.ac.uk}
\author{S. L. Danilishin}
\author{A. Gl\"{a}fke}
\author{B. W. Barr}
\author{A. S. Bell}
\author{C. Gr\"{a}f}
\author{J.-S. Hennig}
\author{E. A. Houston}
\author{S. H. Huttner}
\affiliation{SUPA, School of Physics and Astronomy, The University of Glasgow, Glasgow, G12\,8QQ, UK}
\author{H. L\"{u}ck}
\affiliation{Albert-Einstein-Institut, Max-Planck-Institut f\"{u}r Gravitationsphysik, D-30167 Hannover, Germany}
\affiliation{Leibniz Universit\"{a}t Hannover, D-30167 Hannover, Germany}
\author{D. Pascucci}
\affiliation{SUPA, School of Physics and Astronomy, The University of Glasgow, Glasgow, G12\,8QQ, UK}
\author{K. Somiya}
\affiliation{Graduate School of Science and Technology, Tokyo Institute of Technology, 2-12-1 Oh-okayama, Meguro-ku, Tokyo 152-8551, Japan}
\author{B. Sorazu}
\author{A. Spencer}
\author{S. Steinlechner}
\author{K. A. Strain}
\author{J. Wright}
\author{T. Zhang}
\author{S. Hild}
\affiliation{SUPA, School of Physics and Astronomy, The University of Glasgow, Glasgow, G12\,8QQ, UK}

\date{\today}

\begin{abstract}
The \ssm{} interferometer topology can potentially provide enhanced sensitivity to gravitational waves in the audio-band compared to equivalent Michelson interferometers. A challenge with the \ssm{} interferometer arises from the intrinsic lack of sensitivity at low frequencies where the velocity-proportional signal is smaller than the noise associated with the sensing of the signal. Using as an example the on-going proof-of-concept \ssm{} experiment in Glasgow, we quantify the problem and present a solution involving the extraction of a small displacement-proportional signal. This displacement signal can be combined with the existing velocity signal to enhance low frequency sensitivity, and we derive optimal filters to accomplish this for different signal strengths. We show that the extraction of the displacement signal for low frequency control purposes can be performed without reducing significantly the quantum non-demolition character of this type of interferometer.
\end{abstract}

\maketitle

\section{\label{sec:intro}Introduction}

  The Advanced LIGO gravitational wave observatories \cite{aligo2015} made the first detection of gravitational waves from a binary black hole in 2015 \cite{Abbott2016} at the start of their joint observing run. In the coming years Advanced Virgo \cite{avirgo2015} in Italy and KAGRA \cite{kagra2013} in Japan are expected to join the network, opening up a new window through which to observe the universe. These detectors are, fundamentally, Michelson interferometers with the addition of \fp{} arm cavities and other optical modifications \cite{aligo2015, aligodetectors2016} to achieve exquisite sensitivity to motion of the test masses. Using the Local Lorenz gauge, an incident gravitational wave can be pictured as a change in length between the cavity mirrors in each arm of the interferometer. The primary degree of freedom a gravitational wave would excite is the differential mode of the cavity lengths, \lminus{}, which can be defined in terms of the position of the end test masses (ETMs) $x_{\textrm{A}}^{\textrm{ETM}}$ and $x_{\textrm{B}}^{\textrm{ETM}}$ in cavities A and B, respectively, as:
  \begin{equation}
    \label{eq:darm}
    \textrm{\lminus{}} = \frac{x_{\textrm{A}}^{\textrm{ETM}} - x_{\textrm{B}}^{\textrm{ETM}}}{2}.
  \end{equation}
  
  Quantum noise, the combination of quantum radiation pressure noise at low frequencies and quantum shot noise at high frequencies, restricts the ultimate sensitivity of classical Michelson interferometers at what is termed the \emph{Standard Quantum Limit} (SQL) \cite{Braginsky1995}. As quantum noise is expected to limit the sensitivity of the aforementioned \emph{second generation} detectors over a wide frequency band, particular effort is being paid to its reduction in proposals for \emph{third-generation} instruments such as the Einstein Telescope facility \cite{hild2011} and LIGO Cosmic Explorer \cite{aligoinst2015}.

  Reduction of quantum noise below the SQL in a particular optical configuration involves a so-called \emph{quantum non-demolition} (QND) measurement \cite{Braginsky1995}: either the use of quantum correlations in the outgoing light or the enhancement of the test mass dynamics by means of light \cite{Chen2011}. Quantum correlations arise naturally between the otherwise uncorrelated sources of quantum noise due to the interaction between the light and the motional degrees of freedom of the interferometer and can be used to effectively avoid radiation pressure noise contributions to the readout signal. This group of methods includes frequency-dependent squeezing \cite{Kimble2001}, variational readout \cite{Kimble2001, Vyatchanin1995, Vyatchanin1996} and speed meters \cite{Braginsky1990, Braginsky2000, Chen2003, Danilishin2004}. The latter approach implies the use of light-mirror interactions to create new test mass dynamics in order to increase their response to the gravitational wave action. This includes so-called ``optical springs'' \cite{Braginsky1999, Buonanno2002, Corbitt2007, Rehbein2008, Gordon2015}, ``optical inertia'' \cite{Khalili2011, Voronchev2012} and ``intracavity schemes'' \cite{Braginsky1997, Khalili2002, Danilishin2006}.
  
  The method on which we focus employs an interferometer topology intrinsically sensitive to a QND observable \cite{Danilishin2012}. The measurement of velocity, itself approximately proportional to the QND observable of the free test mass momentum, is one way in which a reduction in quantum radiation pressure noise can be achieved. A proof-of-concept experiment is under way at the University of Glasgow to demonstrate an audio-band reduction of quantum radiation pressure noise in a \ssm{} topology over an equivalent Michelson design \cite{Graef2014}. This topology is being considered as an alternative to the Einstein Telescope's Michelson interferometer design \cite{MuellerEbhardt2009a, Voronchev2015}. The topology under investigation in the proof-of-concept experiment utilises a zero-area Sagnac interferometer with the addition of arm cavities based on the concept presented in ref.\,\cite{Chen2003}. By design, this interferometer produces a signal at the beam splitter's output port proportional to differential arm cavity mirror \emph{velocity}, in contrast to the \emph{displacement}-proportional signal sensed in the Michelson topology.
  
  The presence of arm cavities within the proof-of-concept \ssm{} gives rise to challenges not previously encountered in the control of gravitational wave detectors and other experiments involving Michelson or Sagnac interferometers, and this aspect will be addressed in the following work. In Section\,\ref{sec:velocity-control} we describe in more detail the proof-of-concept \ssm{} experiment and its control requirements. We then describe a control strategy for the \ssm{}'s differential degree of freedom based on that of Michelson designs, and demonstrate the control challenges this approach introduces. In Section\,\ref{sec:mixed-control} we present an alternative strategy which achieves adequate control of the interferometer to reach its design sensitivity over extended periods, and in Section\,\ref{sec:noise-budget} we present a noise budget of the \ssm{} using the alternative control strategy. A summary is provided in Section\,\ref{sec:summary}.

\section{\label{sec:velocity-control}Velocity control}
  
\begin{figure}
    \subfigure[]{\includegraphics[width=0.9\columnwidth]{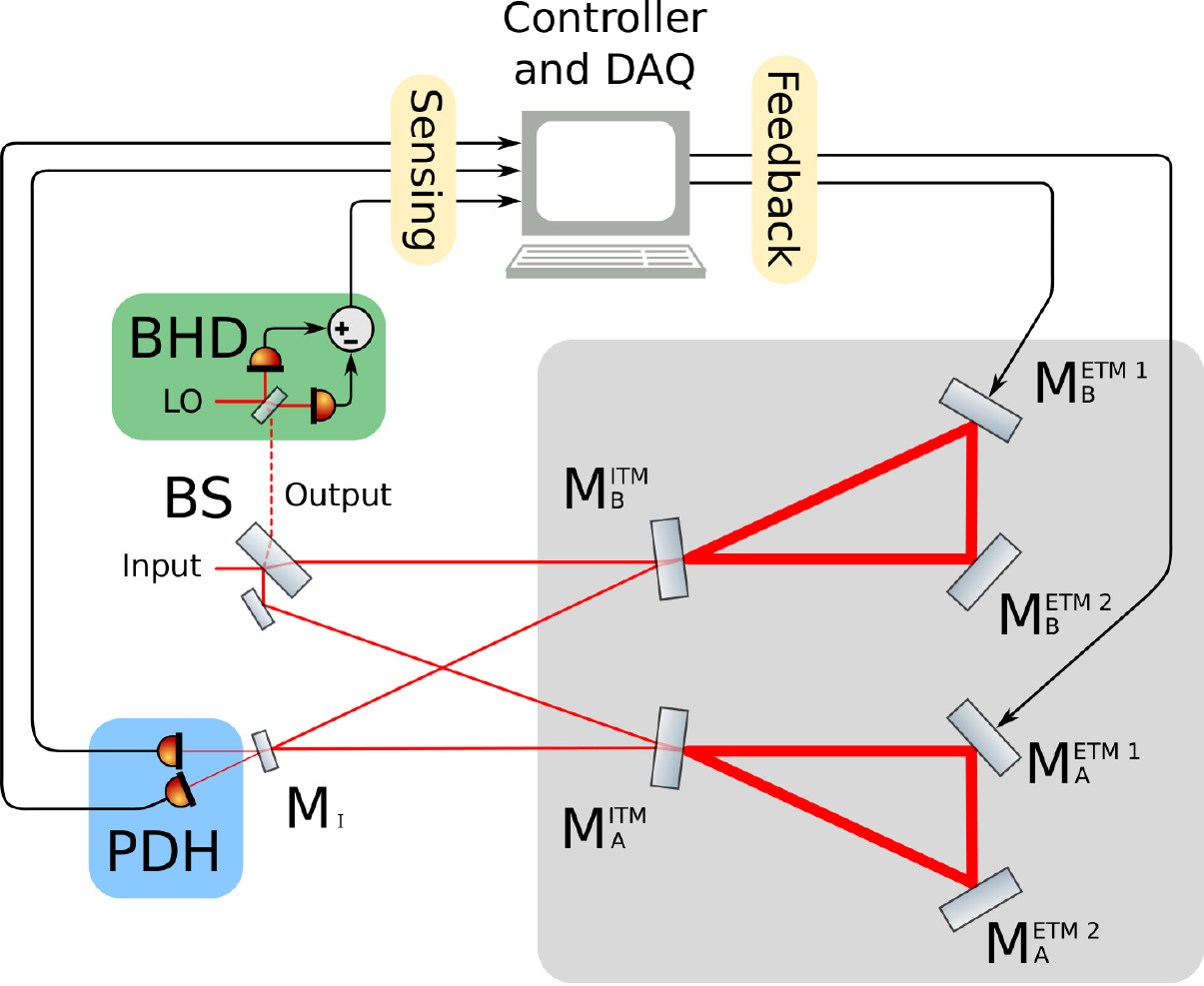}\label{fig:speedmeter-layout}}
    \subfigure[]{\includegraphics[width=\columnwidth]{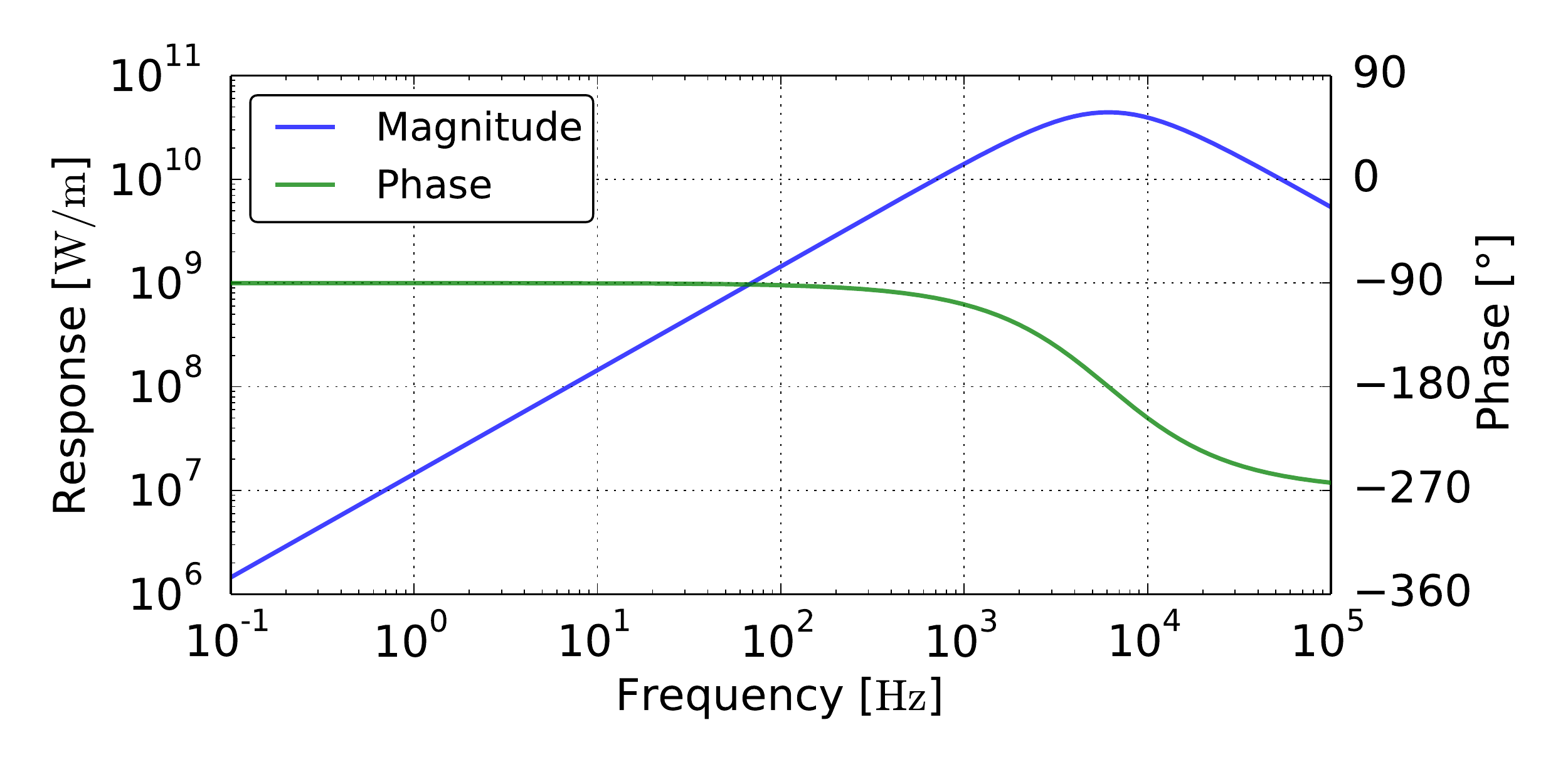}\label{fig:bhd-response}}
    \subfigure[]{\includegraphics[width=0.8\columnwidth]{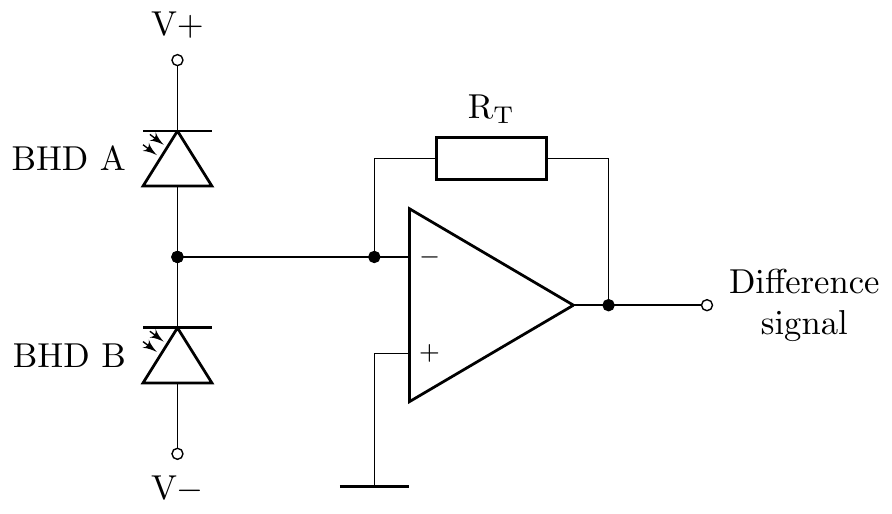}\label{fig:bhd-electronics}}
    \caption{\ssm{} optical layout and extraction of the BHD signal sensitive to the arm cavity differential mode. Figure\,(a) contains a simplified layout of the experiment. Light from the input optics (not shown) is incident upon the main beam splitter. The shaded green area shows the BHD extracting the signal from the main beam splitter's output port (see Section\,\ref{sec:bhd}). The triangular arm cavities are shown in the shaded grey area, and mirror \mi{} couples light between them. Two additional photodetectors (shaded blue area) are present behind \mi{} for individual arm cavity signal extraction. The sensing and feedback signal paths are described in detail in Sections\,\ref{sec:velocity-control} and \ref{sec:mixed-control}. Figure\,(b) contains the frequency response of the BHD to \lminus{}, simulated numerically with \emph{Optickle}. Figure\,(c) contains a simplified version of the intended balanced homodyne detector readout electronics. The difference current from two matched, high quantum efficiency photodetectors is amplified via a transimpedance op-amp stage, with this signal representing the differential velocity of the arm cavity mirrors (see Equation\,\ref{eq:asymdarmbhdresponse}).}
  \end{figure}

  Figure\,\ref{fig:speedmeter-layout} shows a simplified optical layout of the proof-of-concept \ssm{}. The main beam splitter (BS) splits the input field towards the two triangular arm cavities where they form counter-propagating modes. One mode from each arm cavity is coupled into the other via the inter-cavity mirror \mi{}, and the other modes recombine at the main beam splitter.
  
  Each arm cavity in the \ssm{} is an independent degree of freedom and so changes to \lminus{} lead to frequency-dependent signals at the output port. Motion of an arm cavity mirror imparts signal sidebands upon the counter-propagating modes. These modes have different optical path lengths to the beam splitter and so the signal at the output port contains the superposition of signals representing the mirror's displacement from different points in time, which is analogous to velocity. At dc frequencies the two modes at the output port contain the same displacement information and the velocity signal is therefore zero (for a more complete description of the \ssm{}'s behaviour, see, for example, Section\,IIb of \cite{Chen2003}). Motion of \mi{} and BS imprints common phase changes on both counter-propagating modes and are not considered degrees of freedom for signals sensed at the output port.
  
  The error signal (\emph{readout}) representing \lminus{} is sensed at the main beam splitter's output port by means of a balanced homodyne detector (BHD, see Section\,\ref{sec:bhd}) \cite{Steinlechner2015}, as shown in the shaded green area in Figure\,\ref{fig:speedmeter-layout}. The frequency dependence of the phase quadrature signal at the output port $s_{\textrm{BHD}}$ is given by the following relationship, ignoring the effect of losses (a full treatment of the effect of loss is given in ref.\,\cite{Danilishin2015}):
  \begin{equation}
    \label{eq:asymdarmbhdresponse}
    s_{\textrm{BHD}} \left( \Omega \right) \propto \frac{\Omega}{ \left(\Omega^2 + \gamma_{\textrm{arm}}^2 \right)} L_{\left(-\right)},
  \end{equation}
  for angular frequency $\Omega$ and with arm cavity half-bandwidth $\gamma_{\textrm{arm}}$ defined to be:
  \begin{equation}
    \gamma_{\textrm{arm}} = \frac{c_{0} T_{\textrm{ITM}}}{4 L_{\textrm{RT}}},
  \end{equation}
  for speed of light $c_{0}$, arm cavity input test mass (ITM) power transmissivity $T_{\textrm{ITM}}$ and arm cavity round-trip length $L_{\textrm{RT}}$.
  
  Other terms in the response function dependent upon mirror mass, laser power and mechanical modes are not frequency dependent. Note that for $\Omega \ll \gamma_{\textrm{arm}}$ the response is proportional to frequency, vanishing towards dc, as described above and shown in Figure\,\ref{fig:bhd-response}.
  
  In order to maintain peak BHD sensitivity to \lminus{} and therefore gravitational waves, the positions of the cavity mirrors are controlled using \emph{linear inverting feedback}, where an error signal is extracted and applied through a control law to cavity mirror actuators. In the experiment, voice coils and plate capacitor electrostatic drives (ESDs) \cite{Wittel2015} are used to actuate on the position of the end test masses within each cavity. This feedback maintains the interferometer close to its operating point within the bandwidth of the controller. In order to achieve the required stability, the relative position of the cavity mirrors must be controlled to within \SI{3.5e-13}{\meter} rms (see Appendix\,\ref{app:required-control}).
  
  
  \subsection{\label{sec:bhd}Balanced homodyne detection}
    
    The BHD consists of two high quantum-efficiency photodetectors sensing the reflected and transmitted fields from the BHD's beam splitter. A local oscillator is incident upon the BHD's beam splitter to provide gain for the velocity information encoded within the light from the main beam splitter. The difference current is converted to a voltage by an op-amp with transimpedance resistor \rt{} before being sent to the data acquisition system (DAQ). An example circuit for the balanced homodyne detector is shown in Figure\,\ref{fig:bhd-electronics}. The op-amp introduces its own noise to the output, though a well-chosen op-amp will possess noise significantly lower than the signal representing \lminus{} in the intended measurement band. In order for an op-amp to contribute less than \SI{1}{\percent} of the uncorrelated noise in the measurement, its noise must be at least a factor of \SI{10}{} below the dominating noise source in the measurement band.
    
    Op-amps used for control in audio-band interferometry typically possess a noise power spectrum inversely proportional to frequency (so-called \emph{flicker noise} \cite[Section\,11.2.3]{Gray2009}) in the low audio band. As the BHD error signal is dependent upon the time derivative of the mirror positions, however, there will necessarily be frequencies at which the op-amp noise will dominate the BHD error signal. This makes control of slow drifts of the arm cavity mirror positions impossible with the velocity readout, despite the op-amp being well-chosen for a measurement band above \SI{100}{\hertz}. This control problem with relation to the proof-of-concept experiment will be quantified in the following subsections.

  \subsection{\label{sec:op-amp-noise}Op-amp noise}
    
    \begin{figure}
      \subfigure[]{\includegraphics[width=0.8\columnwidth]{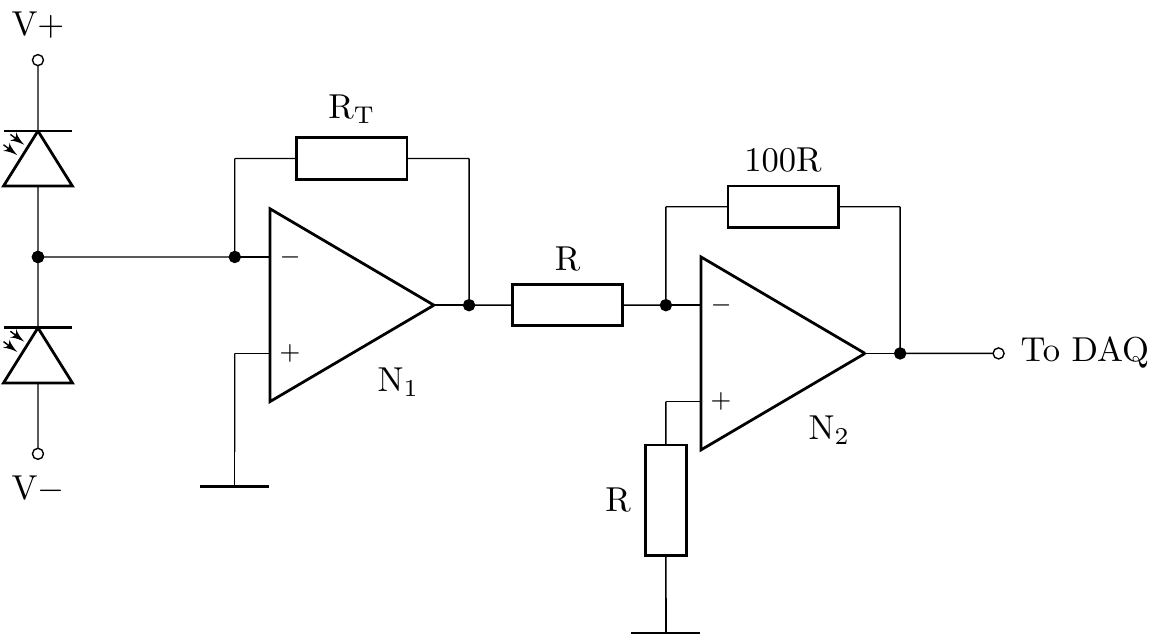}\label{fig:bhd-noise-electronics}}
      \subfigure[]{\includegraphics[width=\columnwidth]{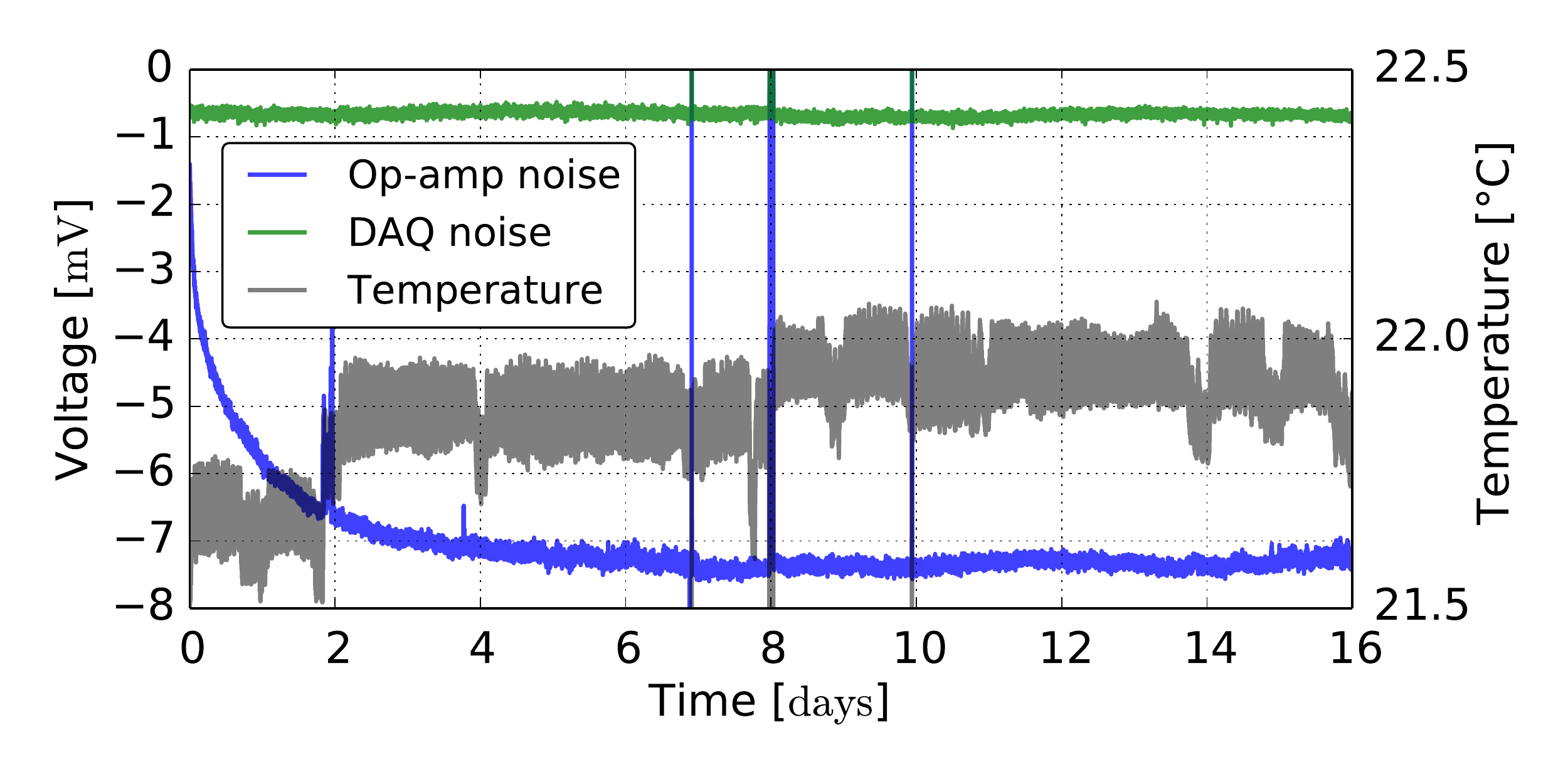}\label{fig:op-amp-noise-time-series}}
      \subfigure[]{\includegraphics[width=\columnwidth]{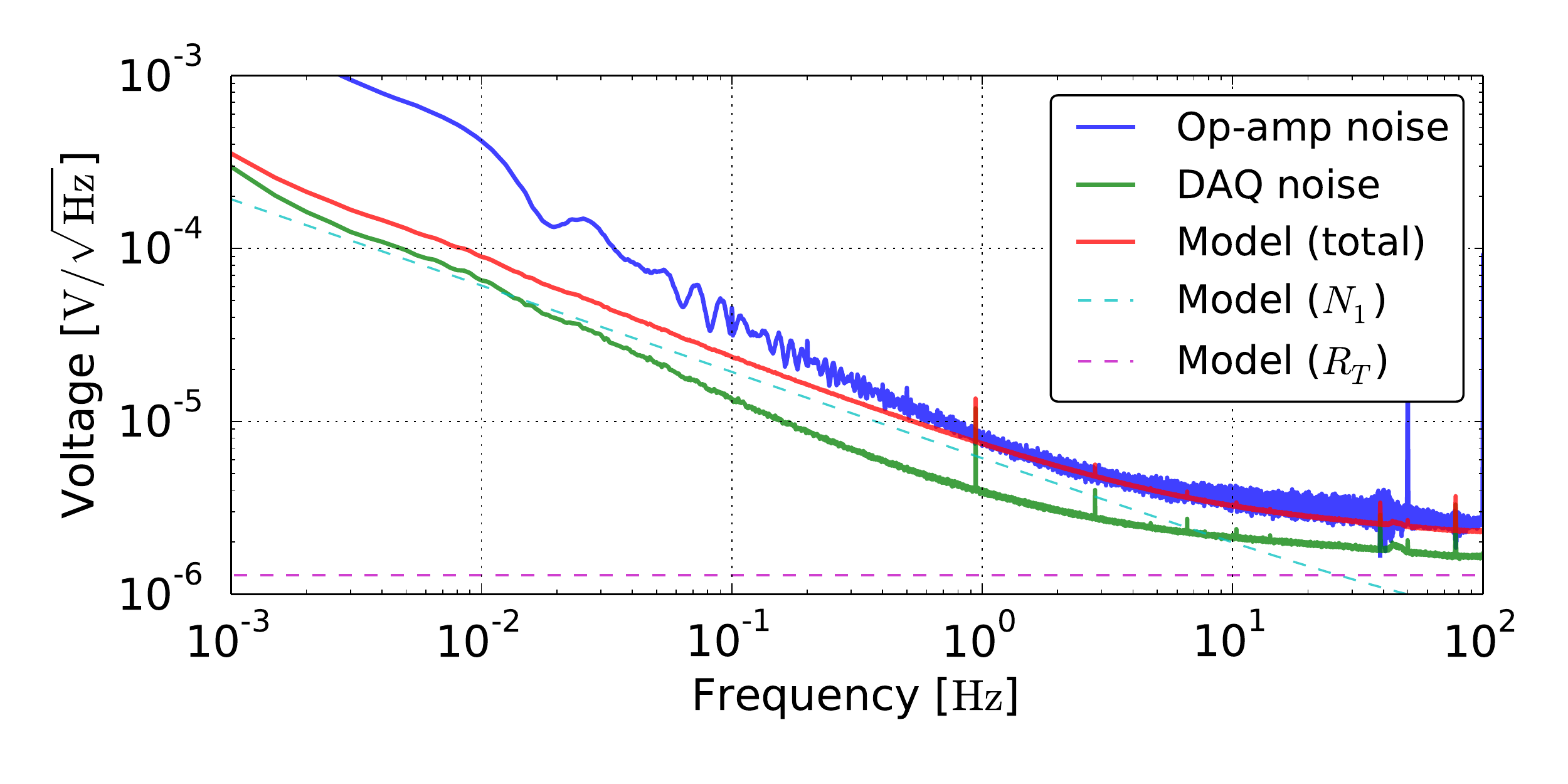}\label{fig:op-amp-noise-spectrum}}
      \caption{Balanced homodyne detector op-amp noise measurements. The electronic schematic is shown in (a). The time series over a period of 16 days is shown in (b) for the op-amp noise at $\textrm{N}_{\textrm{1}}$ (blue) along with the DAQ noise for an empty channel (green). Temperature measurements (black) are shown on the second (right) vertical axis. A drift in the op-amp noise of around \SI{7}{\milli\volt} is visible in the first week. The spectral density in (c) shows the data in (b) plotted in the frequency domain using a Fourier transform with windows of \SI{e4}{\second}. The op-amp and DAQ noise spectra are shown along with modelled op-amp and resistor noise sources projected into the same measurement point. The op-amp noise resembles the modelled total noise (red) at frequencies above \SI{1}{\hertz}.}
    \end{figure}
    
    To measure the effect of a suitable op-amp's noise at low frequency, the output from an applicable BHD circuit was investigated. The circuit shown in Figure\,\ref{fig:bhd-noise-electronics} was housed within a dark enclosure to minimise photocurrent, with one of the two op-amps within a Texas Instruments\textsuperscript{\textregistered} OPA2227 integrated circuit being used to amplify the noise from the other by a factor of \SI{100}{}, to a level detectable by the DAQ. This op-amp is suitable for the BHD circuit shown in Figure\,\ref{fig:bhd-electronics} given the intended measurement band around a few hundreds of \SI{}{\hertz} \cite{Graef2014}.

    The time series in Figure\,\ref{fig:op-amp-noise-time-series} shows a drift in the measured op-amp noise over the course of \SI{16}{} days. An open channel on the DAQ was measured concurrently. A Fourier transform of the measured op-amp noise time series (Figure\,\ref{fig:op-amp-noise-spectrum}) shows a combination of flicker noise and an additional slope possibly due to resistor current noise below around \SI{1}{\hertz} \cite{Seifert2009}. DAQ noise dominates above \SI{4}{\hertz}. The ``Model (total)'' spectral density in Figure\,\ref{fig:op-amp-noise-spectrum} show the contributions to the measurements from the first op-amp $\textrm{N}_{1}$'s current and voltage noise and the Johnson-Nyquist noise of its transimpedance resistor $\textrm{R}_{\textrm{T}}$. This spectral density additionally contains the measured open channel noise summed in quadrature to show the agreement it has with the measurements down to around \SI{1}{\hertz}.
    
    The op-amp noise drift produces an offset upon the BHD error signal which is to be fed back to the cavity mirror actuators, and thus op-amp noise directly contributes to cavity mirror displacement noise, affecting the experiment's sensitivity to the arm cavity differential mode. Since the signal measured at the BHD represents cavity mirror velocity, it must necessarily drop below the noise at low frequencies where the velocity tends to zero.
    
  \subsection{Noise projection}
  
    To reach the desired sensitivity of the interferometer it is crucial to understand the noise characteristics associated with the sensing and control apparatus employed in the experiment. Individual noise sources, for example arising from the BHD op-amp electronics, can be projected into units of differential displacement-equivalent noise using the linear projection technique \cite{Smith2006}. The sources of noise can be logically separated into two broad categories: \emph{sensing noise} and \emph{displacement noise}. Both sources of noise are fed back to the test masses because in practice it is not possible for the controller to distinguish them.
    
    Sources of sensing noise are associated with the readout of the variable of interest\textemdash in the case of the \ssm{} the positions of the test masses' surfaces\textemdash but do not directly influence the variable of interest in an open loop measurement. Sources of sensing noise include quantum shot noise, electronic noise including op-amp noise as modelled in Section\,\ref{sec:op-amp-noise} and digitisation noise due to the analogue-to-digital converter (ADC).
    
    Displacement noise sources directly influence the positions of the test mass surfaces being measured by the interferometer and are therefore transformed by the dynamics of the test masses \cite{Danilishin2015}. As the readout variable in the BHD is the time derivative of position, the control system measures and actively suppresses these noise sources. Significant sources of displacement noise in the \ssm{} experiment are quantum radiation pressure noise, seismic noise, suspension thermal noise \cite{Hammond2012} and coating brownian noise arising from the dielectric coatings present upon the cavity mirrors \cite{Harry2002}.
    
\begin{figure}
      \subfigure[]{\includegraphics[width=\columnwidth]{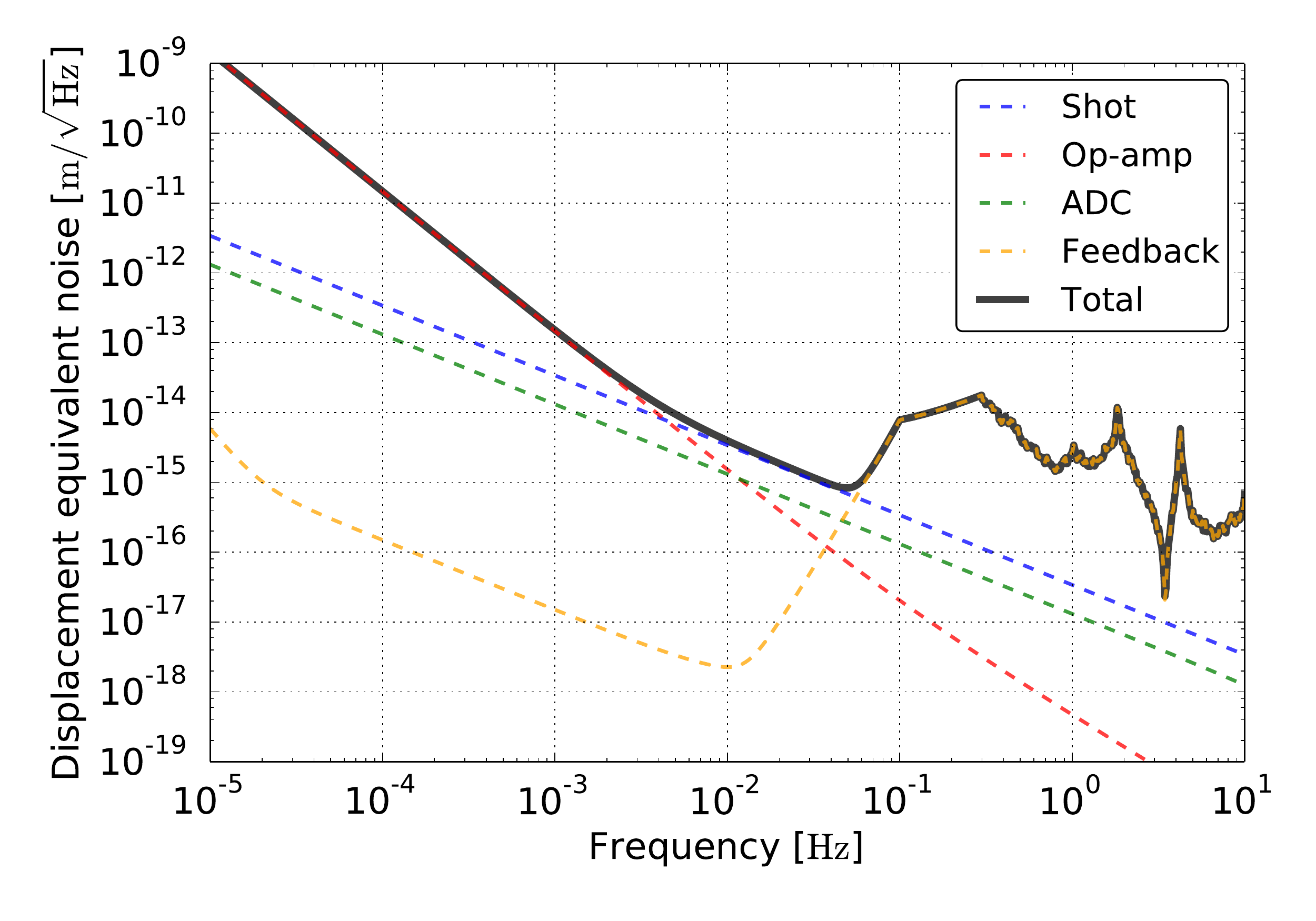}\label{fig:sensing-noise-velocity}}
      \subfigure[]{\includegraphics[width=\columnwidth]{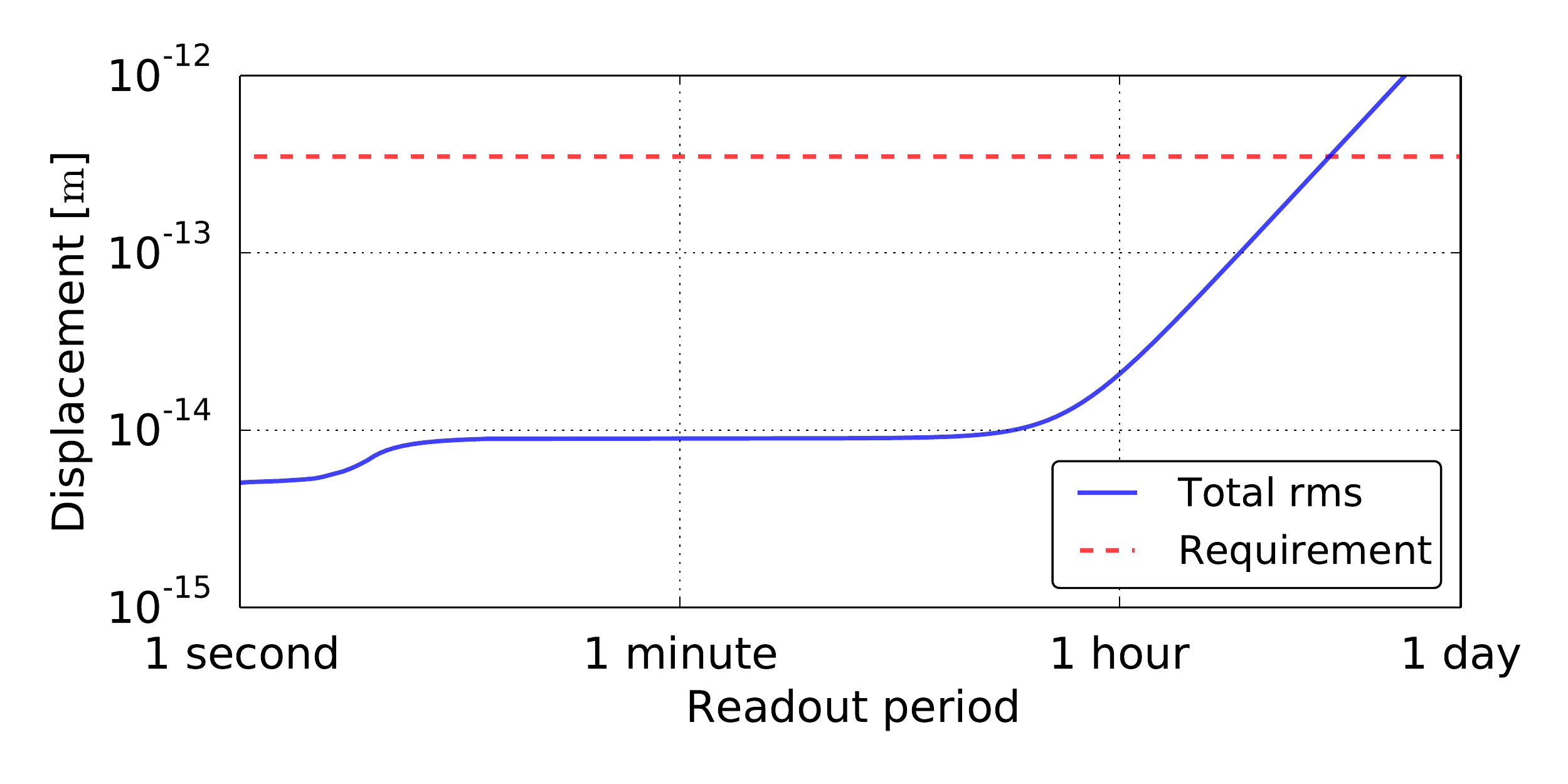}\label{fig:sensing-noise-velocity-rms}}
      \caption{Noise associated with the velocity readout scheme. The spectral density in (a) shows the noise associated with the readout of the signal representing \lminus{} using the BHD. The significant noise sources associated with sensing and feedback are shown. Lab measurements of seismic noise have been made down to \SI{0.3}{\hertz}, and the assumption has been made that the noise is sharply suppressed below the microseism at \SI{0.1}{\hertz}. Below \SI{20}{\milli\hertz} the dominating readout noise is due to the op-amp electronics. The time series in (b) shows the root-mean-square arm cavity mirror motion due to readout noise over one day. Beyond a few hours the motion is greater than the control requirement presented in Appendix\,\ref{app:required-control}}
    \end{figure}

    The noise projection for \lminus{}, calculated using the numerical optomechanical simulation tool \emph{Optickle} \cite{Evans2012} and the control noise modelling tool \emph{SimulinkNb} \cite{SimulinkNb}, is shown in Figure\,\ref{fig:sensing-noise-velocity}. The root-mean-square (rms) differential displacement this creates is shown in Figure\,\ref{fig:sensing-noise-velocity-rms} as a function of time. It shows that, as the interferometer is held at its operating point, over a period of several hours the expected drift is large enough for the cavities to become uncontrollable (see Appendix\,\ref{app:required-control}).
    
    Although for sensing noise we only consider electronic and shot noise, in the experiment there will be other contributing forms of time-varying offset present upon the BHD error signal:
    \begin{itemize}
      \item residual local oscillator light due to temperature-driven imbalances in the BHD beam splitting ratio and photodetector quantum efficiencies;
      \item common mode arm cavity motion due to imbalanced beam splitting at the main beam splitter \cite{Danilishin2015};
      \item thermoelectric potentials and op-amp drift in pre-amplifier and whitening electronics;
      \item any other time-varying effects.
    \end{itemize}
    As such, the estimated rms displacement shown in Figure\,\ref{fig:sensing-noise-velocity-rms} represents a ``best case'' scenario where the op-amp's electronic noise is the dominant effect at low frequencies, and this drift becomes unacceptably large after a few hours. To allow for long term cavity stability it is essential for the error signal to contain a signal significantly above the electronic noise at low frequencies. In the next section we present a strategy for obtaining an error signal of suitable magnitude across the entire control bandwidth.

\section{\label{sec:mixed-control}Velocity-displacement control}
  
  Light from each counter-propagating mode is incident upon the inter-cavity steering mirror \mi{}, and as such this is a natural port in which to separate the modes and sense the motion of each arm cavity (see the shaded blue region of Figure\,\ref{fig:speedmeter-layout}). Using RF modulation, for instance \emph{via} the Pound-Drever-Hall (PDH) technique \cite{Drever1983}, it is possible to obtain a displacement error signal for each cavity that, unlike the velocity signal from the BHD, has flat response at dc, with a similar cavity pole frequency. The individual cavity PDH signals can be mixed to obtain a measurement of \lminus{}, and the frequency dependence of the signal $s_{\textrm{PDH}}$ is, following ref.\,\cite{Kimble2001}, given by:    
  \begin{equation}
    \label{eq:m9darmpdhresponse}
    s_{\textrm{PDH}} \left( \Omega \right) \propto \sqrt{\frac{\gamma_{\textrm{arm}}}{\left(\Omega^2 + \gamma_{\textrm{arm}}^2 \right)}} L_{\left(-\right)},
  \end{equation}
  ignoring again the effect of losses and constant terms as with Equation \ref{eq:asymdarmbhdresponse}. Note that for $\Omega \ll \gamma_{\textrm{arm}}$, the response is flat as expected for a displacement measurement and as such the PDH readout offers a suitable signal to sense \lminus{} at low frequencies.

  \subsection{\label{sec:combined-filter}Combined filter}
  
    The separate velocity and displacement readouts contain the same fundamental information about the position of the mirrors, albeit with different response functions. We can express the signal at output field $i$ as a function of the $k^{\textrm{th}}$ mode of motion, $\hat{o}_{\textrm{k,i}} \left( \Omega \right)$, as \cite{Kimble2001}:
    \begin{equation}
      \label{eq:readout-signals}
      \hat{o}_{\textrm{k,i}} \left( \Omega \right) = L_{\textrm{k}}\left(\Omega\right) + \frac{\hat{n}_{\textrm{i}} \left( \Omega \right)}{R_{\textrm{k,i}} \left( \Omega \right)}
    \end{equation}
    where $L_{\textrm{k}}$ is the position of mode $k$, $\hat{n}_{\textrm{i}} \left( \Omega \right)$ is the noise at field $i$ and $R_{\textrm{k,i}} \left( \Omega \right)$ is the optomechanical transfer function of mode $k$ to field $i$. The definition of a field in this case refers to that of a single signal sideband, $\Omega$. The total time domain signal on a perfect sensor due to the $k^{\textrm{th}}$ mode at the location of the output field will see a combination of the upper and lower signal sidebands:
    \begin{equation}
      \hat{o}_{\textrm{k,i}} \left( t \right) = \int_{0}^{\infty} \frac{\textrm{d} \Omega}{2 \pi} \left( \hat{o}_{\textrm{k,i}} \left( \omega_{0} + \Omega \right) + \hat{o}_{\textrm{k,i}}^\dag \left( \omega_{0} - \Omega \right) \right) e^{-i \Omega t},
    \end{equation}
    where $\omega_{0}$ is the angular frequency of the carrier.
    
    Classical noise sources associated with the test mass modes, such as thermal and seismic noise, are implicit in $L$. The excess noise at each readout port is therefore due to $\hat{n}_{\textrm{i}}$, the quantum vacuum entering at open ports within the interferometer. The presence of such vacuum noise limits the sensitivity of the interferometer in the measurement band. For this reason the reflectivity of \mi{} must be chosen to be close to unity, therefore only a small amount of light is available to the displacement readout for use as a low frequency error signal.
    
\begin{figure}
      \subfigure[]{\includegraphics[width=\columnwidth]{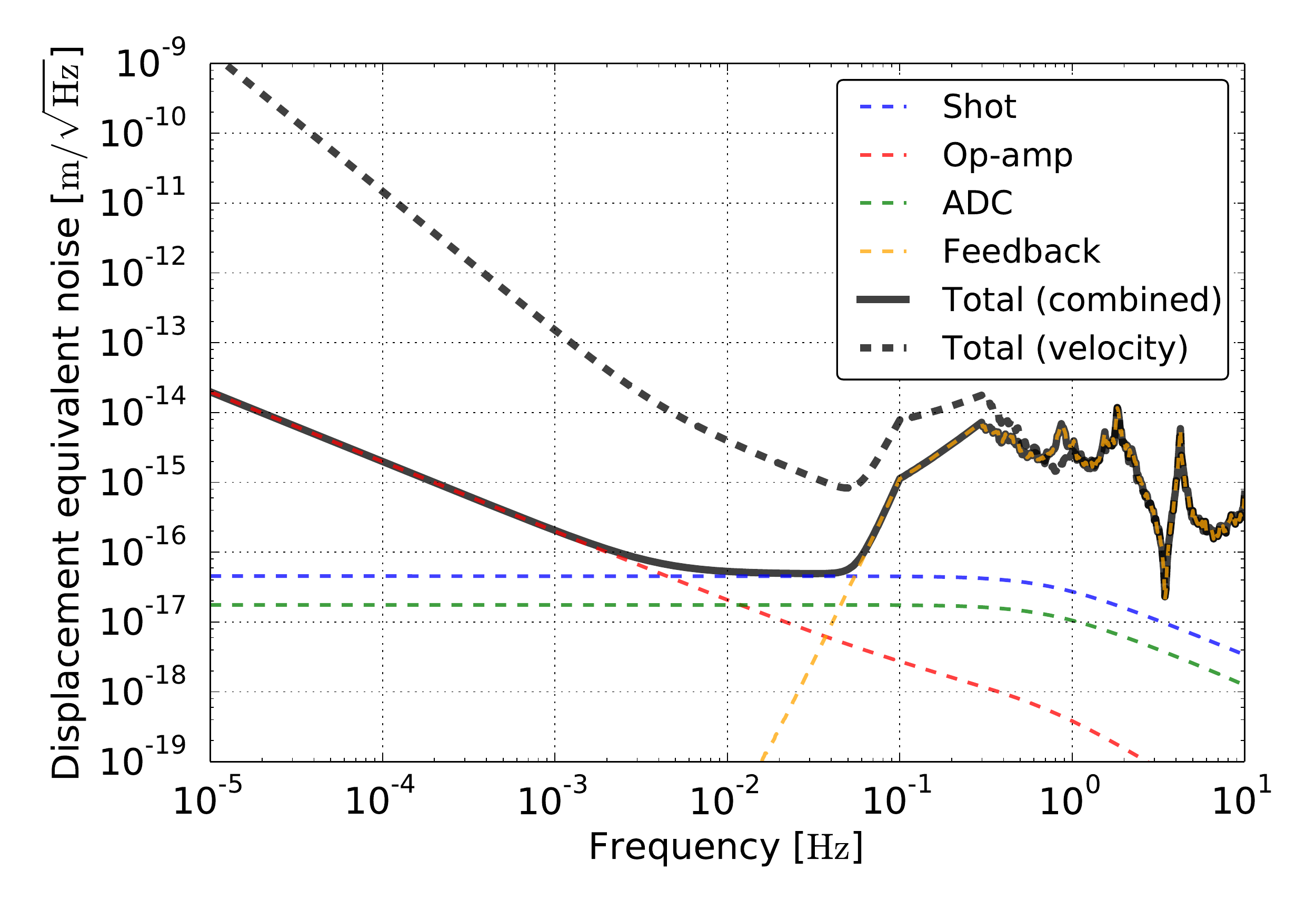}\label{fig:sensing-noise-mixed}}
      \subfigure[]{\includegraphics[width=\columnwidth]{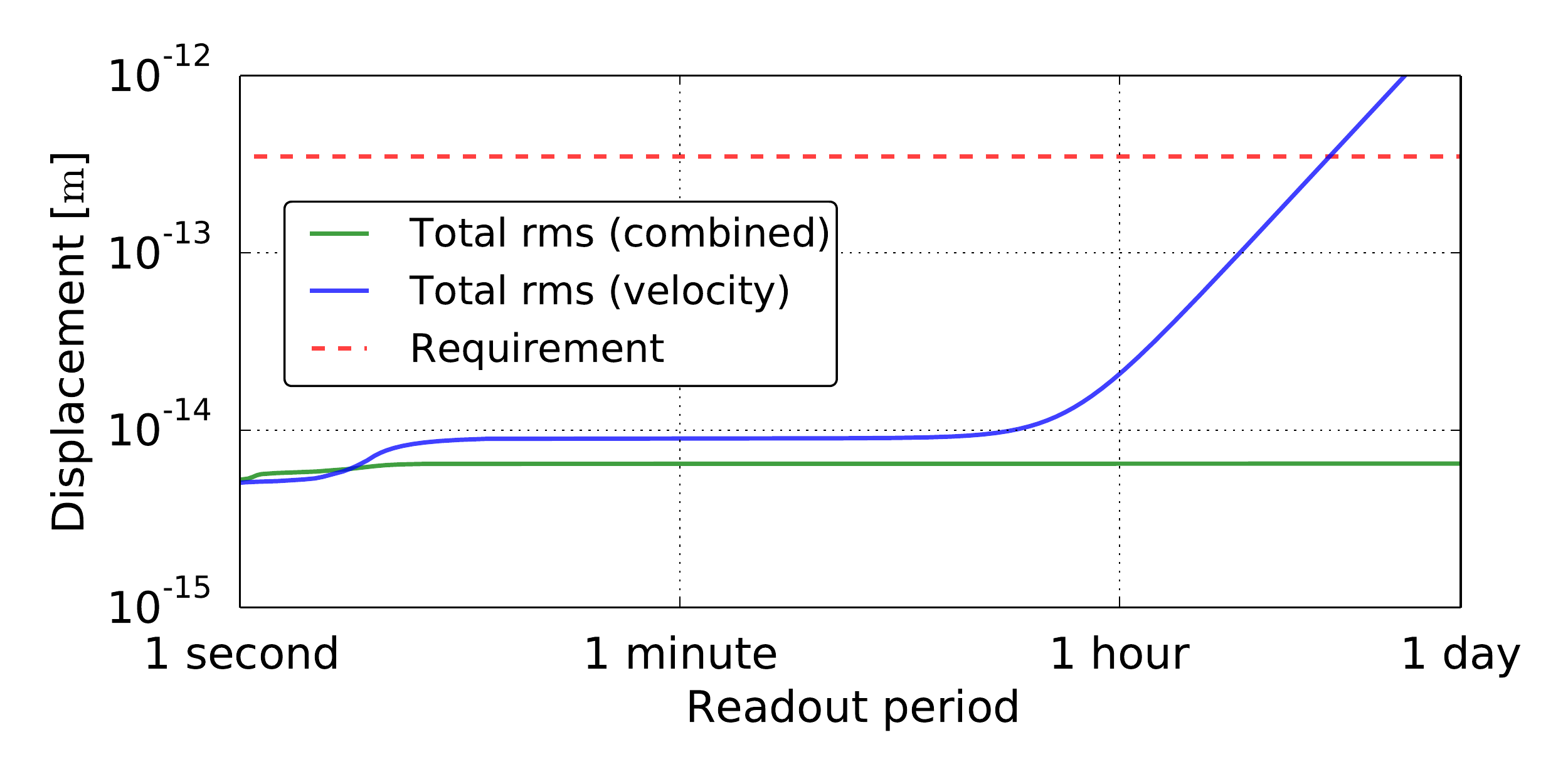}\label{fig:sensing-noise-mixed-rms}}
      \caption{Combined filter readout noise. The spectral density in (a) shows the noise due to sensing and control using a combination of velocity and displacement readout schemes. The total noise in the velocity-only scheme is shown for reference. The mixing of displacement information into the feedback signal at low frequencies leads to greatly reduced displacement noise. The time series in (b) shows the root-mean-square differential displacement the combined filter causes. The combined filter provides a displacement reference at low frequencies and allows control of the \ssm{} for significantly longer periods than with only a velocity readout.}
    \end{figure}

    By considering the response and noise characteristics of the BHD and PDH readouts it is possible to combine them with a filter in order to maximise the interferometer's sensitivity across the full intended frequency range. A desirable crossover frequency for this filter is constrained from below by the signal-to-noise ratio of the BHD and from above by the noise introduced onto the feedback signal by the PDH readout. There is a \SI{90}{\degree} phase difference between the displacement and velocity signals at low frequencies and as such simply combining the PDH and BHD signals with dc gain produces a filter with a stable crossover that can be used as an error signal. The feedback of this combined filter output allows the displacement signal from the PDH to control the cavity mirrors at low frequencies where it is stronger, while letting the BHD signal provide feedback at higher frequencies where it yields the greatest response. The differential displacement-equivalent noise projection for a suitable combined filter is shown in Figure\,\ref{fig:sensing-noise-mixed}. The rms displacement in Figure\,\ref{fig:sensing-noise-mixed-rms} shows a clear reduction in residual displacement with respect to the feedback using only the velocity signal.
    
  \subsection{\label{sec:optimal-filter}Optimal filter}
  
    By considering cross-correlations in the quantum noise at the BHD and PDH readouts, it is possible to produce an optimal filter with which to combine the two in such a way as to minimise the total noise spectral density. The noise at each readout is the sum of the quantum noise inputs at open ports propagated through the interferometer with appropriate transfer functions, so we can rewrite $\hat{n}_{\textrm{i}}$ in Equation\,\ref{eq:readout-signals} in terms of the quantum noise amplitudes $\hat{q}_{\textrm{m}}$ entering at $N_{\textrm{p}}$ open ports:
    \begin{equation}
      \hat{n}_{\textrm{i}} \left( \Omega \right) = \sum_{m=1}^{N_{\textrm{p}}} M^{\textrm{ff}}_{\textrm{m,i}}\left( \Omega \right) \hat{q}_{\textrm{m}} \left( \Omega \right),
    \end{equation}
    where $M^{\textrm{ff}}_{\textrm{m,i}}\left( \Omega \right)$ represents the transfer function between input field $m$ and output field $i$ for signal sideband $\Omega$. The cross-correlation spectral density for unity noise at the $i^{\textrm{th}}$ and $j^{\textrm{th}}$ output channels, for the $k^{\textrm{th}}$ mode, is then \cite{Danilishin2012}:
    \begin{equation}
      \begin{split}
	S_{\textrm{k,\,ij}}(\Omega) = \sum_{m=1}^{N_{\textrm{p}}} \dfrac{\left[M^{\textrm{ff\,*}}_{\textrm{m,\,i}}(\Omega)M^{\textrm{ff}}_{\textrm{m,\,j}}(\Omega)+M^{\textrm{ff\,*}}_{\textrm{m,\,j}}(-\Omega)M^{\textrm{ff}}_{\textrm{m,\,i}}(-\Omega)\right]}{[R^*_{\textrm{k,\,i}}(\Omega)+R_{\textrm{k,\,i}}(-\Omega)][R_{\textrm{k,\,j}}(\Omega)+R^*_{\textrm{k,\,j}}(-\Omega)]}.
      \end{split}
    \end{equation}
    This reduces to the following form for noise entering the same port in which it exits:
    \begin{equation}
      S_{i,i} = \frac{1}{2} \frac{\left| M^{\textrm{ff}}_{i,i}\left( \Omega \right) \right|^{2} + \left| M^{\textrm{ff}*}_{i,i}\left( -\Omega \right) \right|^{2}}{\left(\left| R^{ }_{k,i}\left( \Omega \right) \right| + \left| R^*_{k,i}\left(-\Omega\right)\right|\right)^{2}}.
    \end{equation}
    Assuming a filter $\alpha\left( \Omega \right)$ combines the BHD ($i = 1$) and PDH ($i = 2$) fields, its output for $L_{\textrm{k}}$ would be:
    \begin{equation}
      \begin{split}
	\hat{o}_{\textrm{k,combined}} \left( \Omega \right) &= \alpha\left( \Omega \right) \hat{o}_{\textrm{k,1}} \left( \Omega \right) + \left( 1 - \alpha\left( \Omega \right) \right) \hat{o}_{\textrm{k,2}} \left( \Omega \right) \\
			    &= \left( \alpha\left( \Omega \right) L_{\textrm{k}} \left( \Omega \right) + \left(1 - \alpha\left( \Omega \right) \right) L_{\textrm{k}} \left( \Omega \right) \right) \\
			    &+ \frac{\alpha\left( \Omega \right) \hat{n}_{\textrm{1}}}{R_{\textrm{k,1}}\left(\Omega\right)} + \frac{\left( 1 - \alpha\left( \Omega \right) \right) \hat{n}_{\textrm{2}}}{R_{\textrm{k,2}} \left(\Omega\right)}.
      \end{split}
    \end{equation}      
    The corresponding total noise power spectral density of the combined readout is then:
    \begin{equation}
      \label{eq:readout-spectral-density}
      \begin{split}
	S_{\textrm{readout}} &= \left| \alpha \right|^{2} S_{n_{1},n_{1}} + \left| 1 - \alpha \right|^{2} S_{n_{2},n_{2}} \\
	&+ \Re \left[ \alpha^* \left(1 - \alpha \right) S_{n_{1},n_{2}} \right] \\
	&+ \Re \left[ \alpha^* \left(1 - \alpha \right) S_{n_{2},n_{1}} \right],
      \end{split}
    \end{equation}
    where $S_{n_{1},n_{1}}$ is the noise power spectral density at the BHD port due to vacuum entering at the BHD port, $S_{n_{2},n_{2}}$ is the noise power spectral density at the PDH port due to vacuum entering at the PDH port, and $S_{n_{1},n_{2}}$ and $S_{n_{2},n_{1}}$ are the noise power spectral densities for noise entering at one port and exiting at the other. The optimal filter $\alpha_{\textrm{opt}}$ can be determined by minimising Equation\,\ref{eq:readout-spectral-density} over $\alpha$:
    \begin{equation}
      \label{eq:optimal-filter}
      \alpha_{\textrm{opt}} = \frac{S_{n_{1},n_{2}} - S^*_{n_{1},n_{2}}}{S_{n_{1},n_{1}} + S_{n_{2},n_{2}} - \Re \left[ S_{n_{1},n_{2}} \right] - \Re \left[ S_{n_{2},n_{1}} \right]}.
    \end{equation}
    The reflectivity of \mi{} is implicit in both the field-to-field and mode-to-field transfer matrices for each signal sideband, $\mathbf{M}^{\textrm{ff}}$ and $\mathbf{R}$, respectively, and as such $\alpha_{\textrm{opt}}$ depends on the value of \mi{}.
    
\begin{figure}
      \includegraphics[width=\columnwidth]{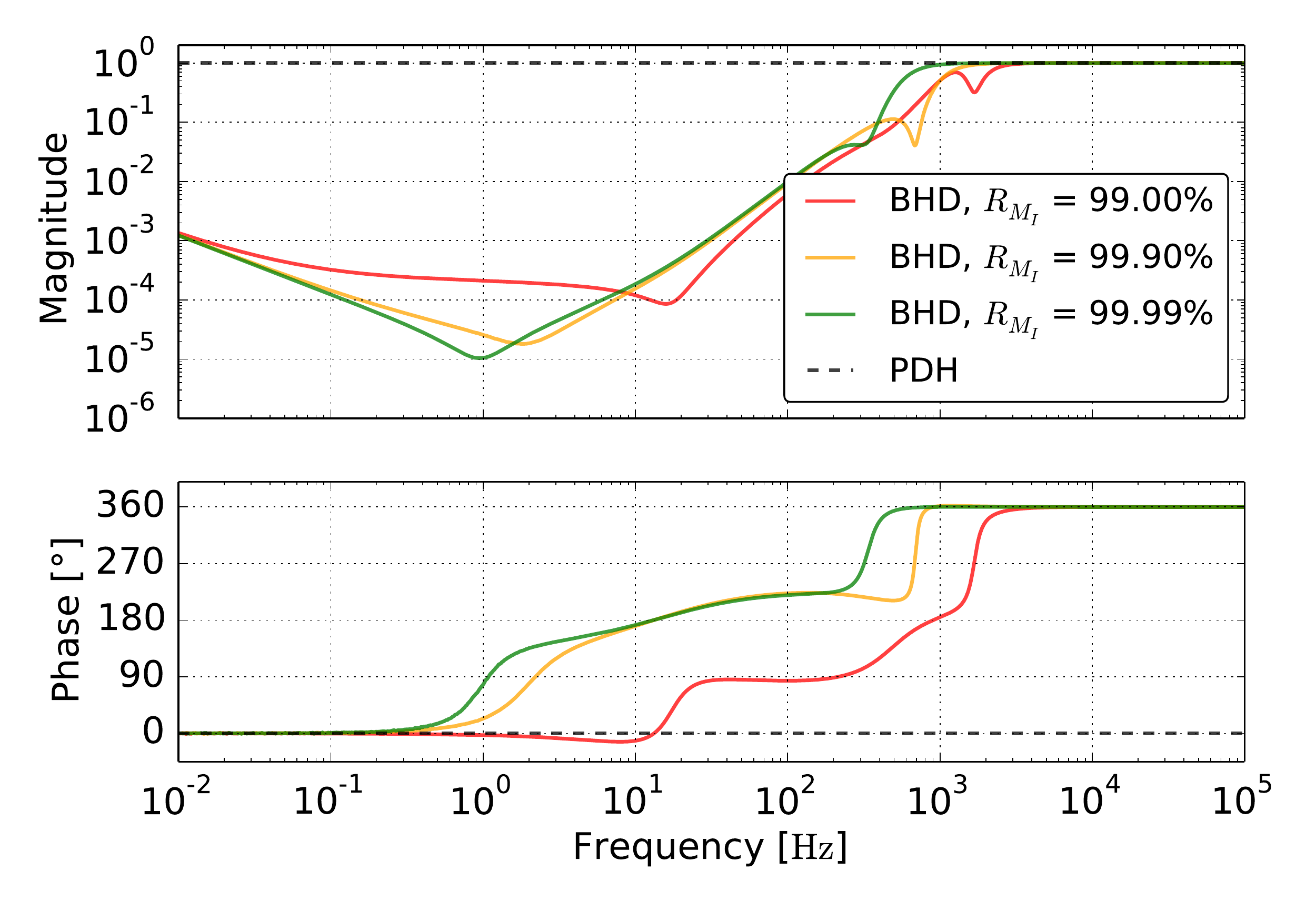}
      \caption{\label{fig:optimal-filters}Optimal filters to combine the BHD and PDH signals for different values of \mi{} reflectivity. The red, yellow and green curves are the coefficients to be applied to the BHD signal with respect to the PDH signal before the two are combined, for different \mi{} (power) reflectivities. The black, dashed curve is the (unity) coefficient to be applied to the PDH signal. For all values of \mi{} shown, the optimal combination involves suppressing the BHD signal with respect to the PDH at frequencies below around \SI{1}{\kilo\hertz}; equivalently, the PDH signal must be amplified with respect to the BHD signal in the same band, an example of which is presented in Section\,\ref{sec:combined-filter}.}
    \end{figure}

    The matrices $\mathbf{M}^{\textrm{ff}}$ and $\mathbf{R}$ are not calculated in Optickle by default, and so some modifications to the code were necessary (see Appendix\,\ref{app:optickle} and \cite{Leavey2016}). The effect of \mi{}'s reflectivity on $\alpha_{\textrm{opt}}$ is shown in Figure\,\ref{fig:optimal-filters}. Note that, because it is calculated with precomputed spectral densities and not tested for stability, the filter predicted by Equation\,\ref{eq:optimal-filter} is not necessarily realisable. A causal Wiener filter has previously been calculated for single-readout interferometers \cite{MuellerEbhardt2009, Miao2010}, but a similar calculation for more than one readout has not yet been investigated. While Equation\,\ref{eq:optimal-filter} enables the lowest noise spectral density for the measurement of the motion of the differential mode of the \ssm{}, in the case of the proof-of-concept experiment simply combining the BHD and PDH signals with dc gain, as suggested in Section\,\ref{sec:combined-filter}, is close to optimal. The difference in response gradients above the cavity pole frequency prevents the PDH signal from contaminating the QND effect in the intended measurement band. The calculation presented in this section, however, is a general solution for any system with multiple readouts for a single variable and may prove useful for future gravitational wave detectors utilising QND techniques.
    
\section{\label{sec:noise-budget}Noise budget}

\begin{figure}
    \includegraphics[width=\columnwidth]{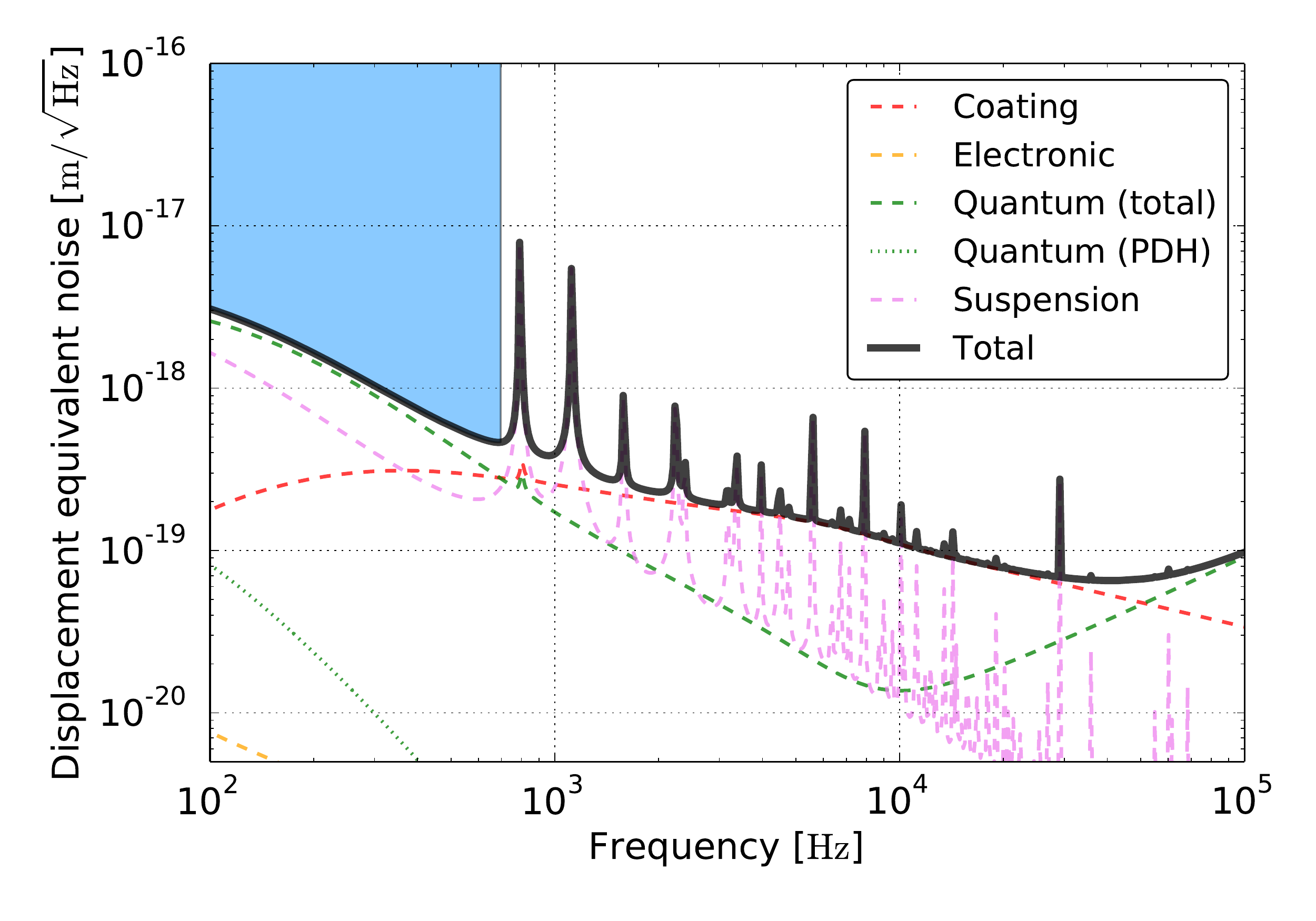}
    \caption{\label{fig:noise-budget}\ssm{} differential mode noise budget for the combined filter scheme with sensing and control noise taken into account. The shaded region represents the frequency band at which the intended direct measurement of reduced quantum radiation pressure noise is to be made in the experiment. The quantum noise contribution from the PDH readout is more than an order of magnitude smaller than the total quantum noise, showing that its inclusion in the combined filter is not harmful to the overall sensitivity in this band.}
  \end{figure}

  In order to show that quantum noise is reduced with respect to an equivalent Michelson interferometer, the design of the proof-of-concept \ssm{} intends for it to be the limiting noise source in a frequency band in the region of a few hundreds of \SI{}{\hertz} \cite{Graef2014}. Using the linear projection technique, each anticipated significant source of noise has been estimated and projected into differential mode displacement-equivalent noise to discover the limiting sources across the control bandwidth, and verify that the experiment will be limited by quantum noise in the intended band. The projection requires that the characteristics of the length sensing and control loop be well defined. A detailed description of the interferometer's optomechanics, signal extraction, control law and suspension and actuator dynamics is provided in Appendix\,\ref{app:control-scheme}, while the computed noise budget for the differential mode is presented in Figure\,\ref{fig:noise-budget}.
  
  The sensitivity between \SI{100}{\hertz} and \SI{700}{\hertz}, shaded in blue, is the quantum noise limited measurement band. This band is constrained from below by test mass suspension mechanical mode cross-couplings (not shown) and from above by the first violin mode of the ETM suspensions. Suspension thermal noise is the second highest noise source present in this band and is at most a factor of \SI{2.3}{} below quantum noise, allowing a careful direct measurement of quantum radiation pressure noise to be made in this region. The contribution to the quantum noise from the PDH feedback is far below the limiting quantum noise, showing that the use of the displacement readout as part of the combined filter presented in Section\,\ref{sec:combined-filter} does not significantly affect the sensitivity of the \ssm{} in the desired band.

\section{\label{sec:summary}Summary}
  
  We have demonstrated that positional drifts of the cavity mirrors in the proof-of-concept \ssm{} at low frequencies due to sensing noise lead to an inability to control the cavity mirrors over time scales longer than a few hours. We have shown that this drift can be suppressed by taking a small amount of light from the path between the arm cavities to provide a displacement readout, and that this does not significantly affect the sensitivity of the main, velocity readout. A combination of the displacement and velocity readouts provides a suitable error signal for the control of the arm cavity differential mode at all relevant frequencies without spoiling the quantum non-demolition effect at higher frequencies, facilitating measurements with arbitrary integration time and allowing the \ssm{} to reach its design sensitivity.
  
  Since the main readout of any interferometer primarily sensitive to velocity will encounter the problem of vanishing signal in the presence of flat or increasing sensing noise at low frequencies, we believe the solution presented in this work is applicable to any audio-band speed-meter.

\begin{acknowledgments}
  The authors would like to thank members of the LIGO, Virgo and KAGRA scientific collaborations for fruitful discussions, and in particular Chris Wipf and Matt Evans. The presented work was made possible by support grants from the European Research Council (ERC-2012-StG: 307245), the Science and Technologies Facility Council (ST/L000946/1 and ST/K502005/1), the International Max Planck Partnership and ET-RD.
  
  This document has been assigned LIGO identifier P1500276.
\end{acknowledgments}

\appendix

\section{\label{app:required-control}Required control}
  
  The noise present within the interferometer will produce an unintended \emph{dark-fringe offset} at the output port of the \ssm{}. Using the parameters listed in Appendix\,\ref{app:parameters} with the relation linking laser frequency fluctuations $\Delta f$ and cavity length fluctuations $\Delta L_{\left(-\right)}$,
  
  \begin{equation}
    \frac{\Delta L_{\left(-\right)}}{L_{\textrm{RT}}} = \frac{\Delta f}{f_{0}},
  \end{equation}
  with $f_{0} = \frac{c_{0}}{\lambda_{0}}$ representing the laser frequency, the requirement for the \ssm{} is that the residual motion of the mirrors must be less than \SI{3.5e-13}{\meter}. This uses the assumption that the frequency fluctuations $\Delta f$ fall within \SI{1}{\percent} of the arm cavity full-width half-maxima\textemdash a common heuristic method used to ensure that technical noise sources do not couple strongly to the gravitational wave channel.
  
  As shown in ref. \cite{Danilishin2015}, main beam splitter asymmetries introduce common arm cavity mode coupling at the output port, which leads to further unintended dark fringe offset, and so the real requirement is likely to be more stringent. The above requirement, however, is sufficient to highlight the problem of velocity-only control as shown in Section\,\ref{sec:velocity-control}.

\section{\label{app:optickle}Calculation of field transfer matrix}

  By default, Optickle will only output the signal and noise on \emph{probes} defined within the system, where a probe is analogous to a photodetector with unity quantum efficiency. A probe signal is a superposition of the field amplitudes in a given location within the interferometer, where the exact linear combination of field amplitudes is determined by the type of probe. In the process of determining a probe signal, the quadrature sum of the field amplitudes immediately in front of the probe is computed. The phase information contained within these fields is lost in this process. Similarly, transfer functions from \emph{drives} (test mass modes within the interferometer) to probes are provided, but not transfer functions from drives to fields.
  
  In order to calculate the cross-correlation spectral density required for the calculation of the optimal filter in Section\,\ref{sec:optimal-filter}, the complex field and drive transfer matrices, $\mathbf{M}^{\textrm{ff}}$ and $\mathbf{R}$, respectively, must be extracted from Optickle indirectly. Optickle's calculation of the quantum noise at each probe within the interferometer uses field-to-field and drive-to-field matrices, but because the quantum noise and drive excitations are not necessarily unity, these matrices are not transfer matrices. In order to obtain $\mathbf{M}^{\textrm{ff}}$ the code which computes the quantum noise at each probe has to be modified to instead inject quantum noise at open ports with unity amplitude. Similarly, $\mathbf{R}$ can be computed by setting the drive amplitudes to unity. The modified source code is publicly available \cite{paperdata}.
  
\section{\label{app:control-scheme}Control scheme}
  
  \begin{figure*}
    \includegraphics[width=0.9\textwidth]{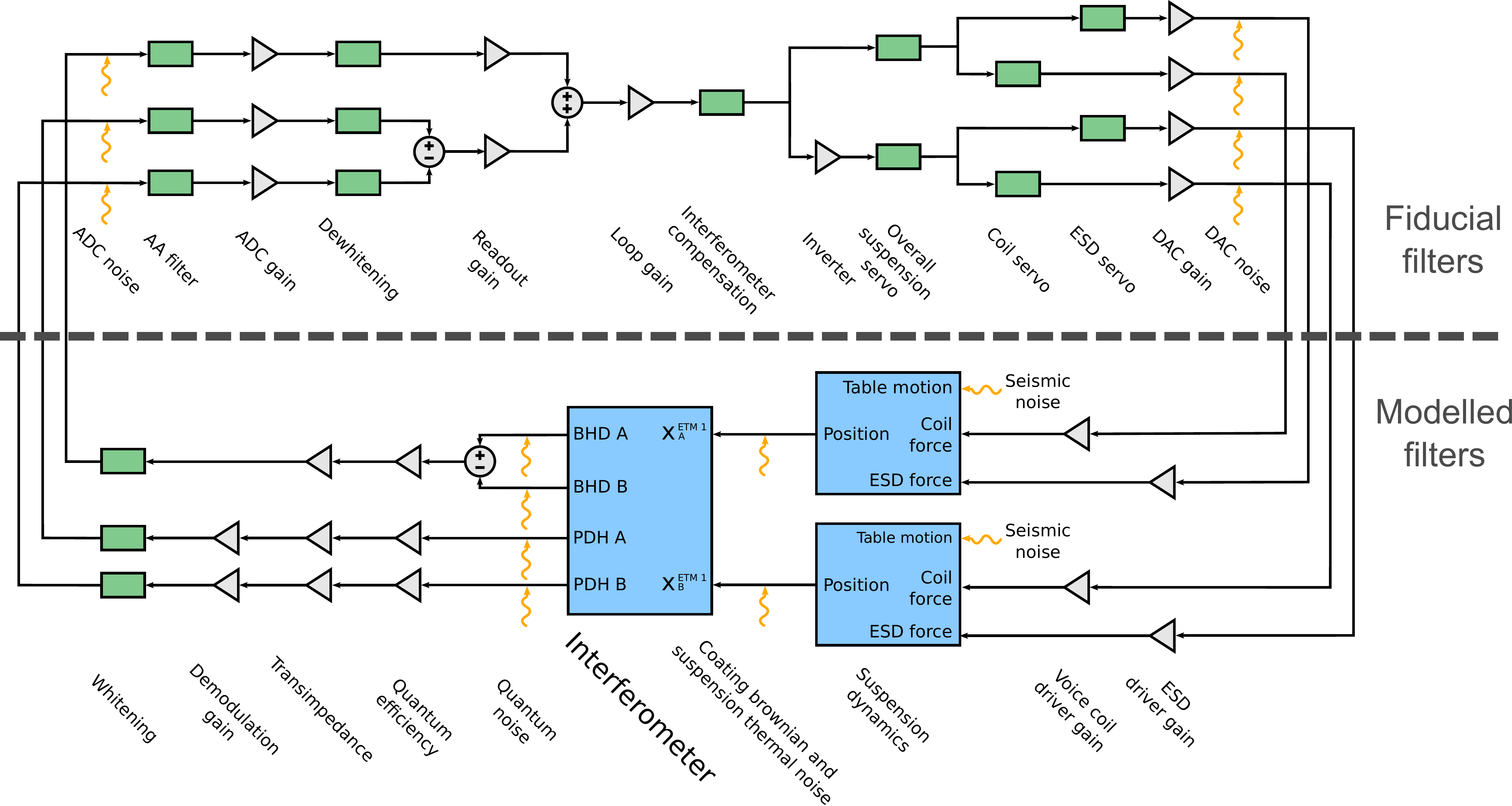}
    \caption{\label{fig:control-loop}\ssm{} control loop model. The interferometer plant produces signals representing the probes in the interferometer, and sensing noise is added before the signals are sent to the digital controller. Within the controller, the interferometer readouts are combined into an error signal representing \lminus{}. The error signal is fed through a series of filters and sent to the test mass actuators, with the addition of DAC noise. The suspension blocks transform the feedback signals into test mass displacements, and seismic, coating brownian and suspension thermal noise is injected at the input to the interferometer plant.}
  \end{figure*}
  
  The intended control loop schematic for the proof-of-concept experiment is shown in Figure\,\ref{fig:control-loop}. The ``fiducial'' section is well defined and is based on existing DAQ hardware and software, an implementation of the LIGO control and data system \cite{Bork2010}. The bottom section contains the blocks which have been modelled during the course of this work.
  
  All blocks shown are frequency dependent matrices with one or more inputs and outputs. The green blocks represent simple filters possessing single inputs and outputs. Each filter multiplies its input by a transfer function in order to calculate its output. The blue blocks represent transfer matrices, which are multiple-input and multiple-output filters. The grey triangles represent dc gain, and are equivalent to simple filters without frequency-dependent response. The undulating orange arrows represent noise injection.
  
  \subsection{Sensing}
  
  The outputs from the interferometer block are probe signals modelled by Optickle for a given set of test mass displacement inputs. Quantum vacuum noise, also modelled by Optickle, is injected onto these signals. The BHD signal is calculated by taking the difference between the two homodyne probe signals, BHD A and BHD B. Each signal is passed through a quantum efficiency gain stage to represent the conversion of light power into photocurrent by the photodetector, and then through a transimpedance gain stage to convert the photocurrent into a voltage. The PDH signals receive additional gain due to RF demodulation. At this point, the signals are sent to the DAQ. Whitening filters ensure the signals are far above the noise of the ADCs. The noise associated with the conversion process is injected at this point. To prevent aliasing due to the ADC sample frequency, \SI{65536}{\hertz}, filters are employed to suppress signal content above around \SI{10}{\kilo\hertz}. The conversion from volts to digital counts within the ADCs is represented by a gain block for each channel.
  
  \subsection{Control}
  
  In the digital domain, dewhitening filters reverse the effect of the whitening filters to recover the original signals. At this point, the arm cavity displacement signal can be constructed by taking the difference between the individual PDH signals. This signal is then combined with the BHD's velocity signal via readout gain stages (as presented in Section\,\ref{sec:combined-filter}) which set the crossover frequency between velocity and displacement measurements. If an optimal filter were to be employed as shown in Section\,\ref{sec:optimal-filter}, these gain blocks would be filters.
  
  The unity gain frequency of the feedback system is set by the loop gain block. A filter is also present here to compensate the feedback signals for the shape of the interferometer's frequency response. The output from this filter is split into two signals to be applied differentially to ETMs $\textrm{M}_{\textrm{A}}^{\textrm{ETM 1}}$ and $\textrm{M}_{\textrm{B}}^{\textrm{ETM 1}}$, as shown in Figure\,\ref{fig:speedmeter-layout}. Actuator range can be further extended by splitting this feedback between the other mirrors in each cavity, but this is not modelled in this work. The signal applied to the $\textrm{M}_{\textrm{B}}^{\textrm{ETM 1}}$ suspension system is inverted in order to move the test masses differentially.
  
  \subsection{Driving}
  
  Each test mass's correction signal is passed through a set of driving filters representing the dynamics of each actuator. The ETMs are suspended from triple pendulums to provide sufficient isolation from seismic noise within the measurement band. In order to provide corrective actuation upon the test masses, voice coil actuators are present on the suspension stage immediately above each test mass in addition to (unsuspended) ESDs situated behind each test mass. The ESDs are sufficiently insensitive to seismic motion transverse to the test mass that they need not be suspended \cite{Wittel2015}. The ESDs' ability to actuate directly upon the test masses makes them suitable for high frequency corrections whereas the extensive range of the voice coils makes them suitable for the suppression of the dominating seismic noise below the measurement band.
  
\begin{figure}
    \includegraphics[width=\columnwidth]{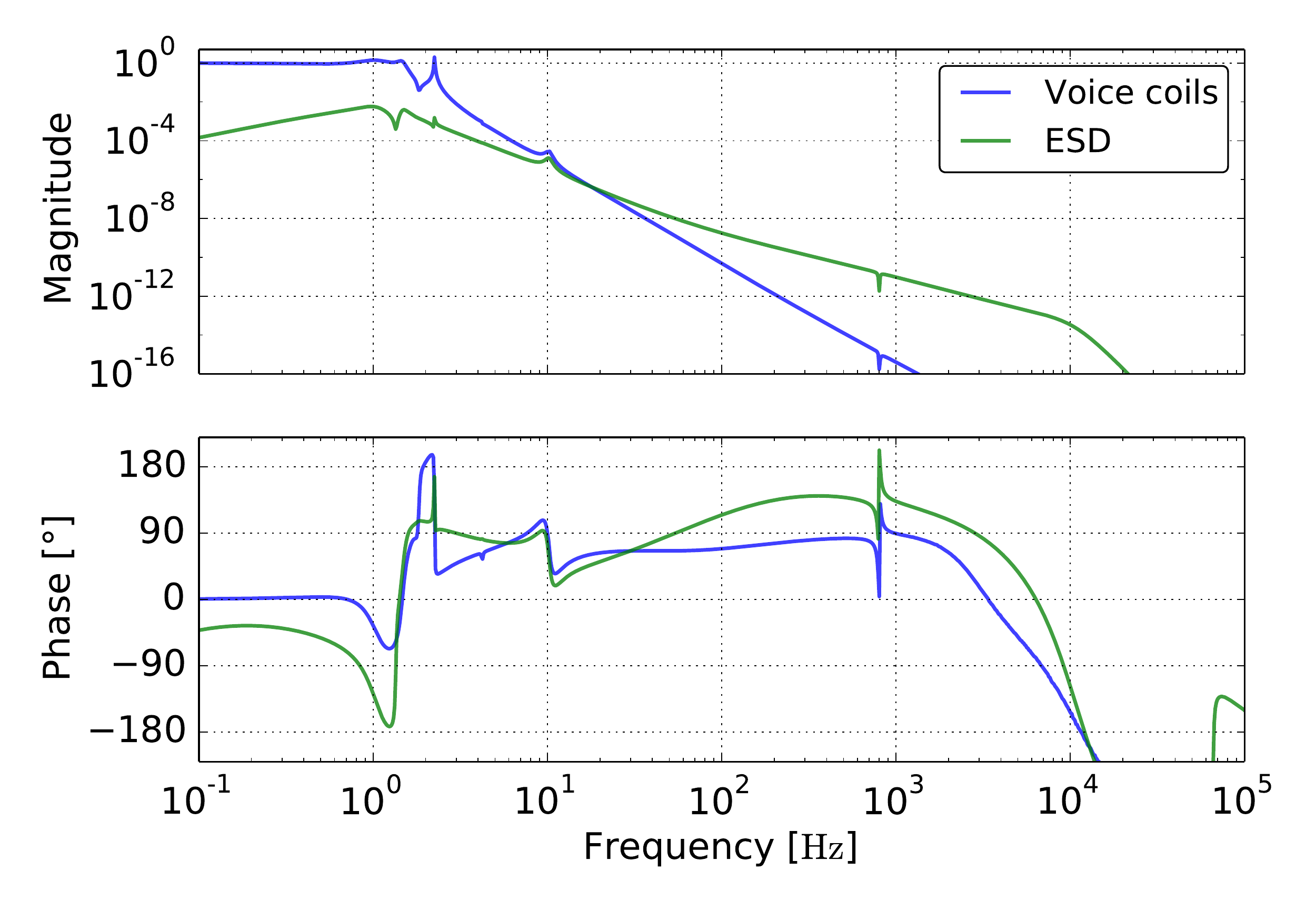}
    \caption{\label{fig:suspension-crossover}Simulated proof-of-principle \ssm{} ETM suspension open loop transfer functions showing the difference in gain between the two test mass actuators. The voice coils provide extensive actuation range but are suppressed at high frequencies by the final stage pendulum. The ESD actuates directly upon the test mass and is therefore capable of providing stronger correction than the voice coils at higher frequencies. Second order low-pass notch filters are present on both actuators at \SI{800}{\hertz} to prevent excitation of the first suspension violin mode.}
  \end{figure}

  Each correction signal is passed through a gain hierarchy to split the feedback between the voice coil and ESD actuators. The open loop transfer functions for each actuator are shown in Figure\,\ref{fig:suspension-crossover}. The crossover frequency between the two actuators is \SI{18}{\hertz}.
  
  Before being sent to the actuators, the driving filter outputs are converted into the analogue domain using appropriate DAC gain and noise injection stages. The conversion from voltage to force is represented by gain stages at the input to each suspension block.
  
  The suspension block contains three inputs, each representing a different passive filtering stage of the suspension system. These input signals are mapped to a single output representing the displacement of each ETM, and these signals are fed back to the interferometer block, completing the loop.
  
  \subsection{Loop Gain}
  
\begin{figure}
    \includegraphics[width=\columnwidth]{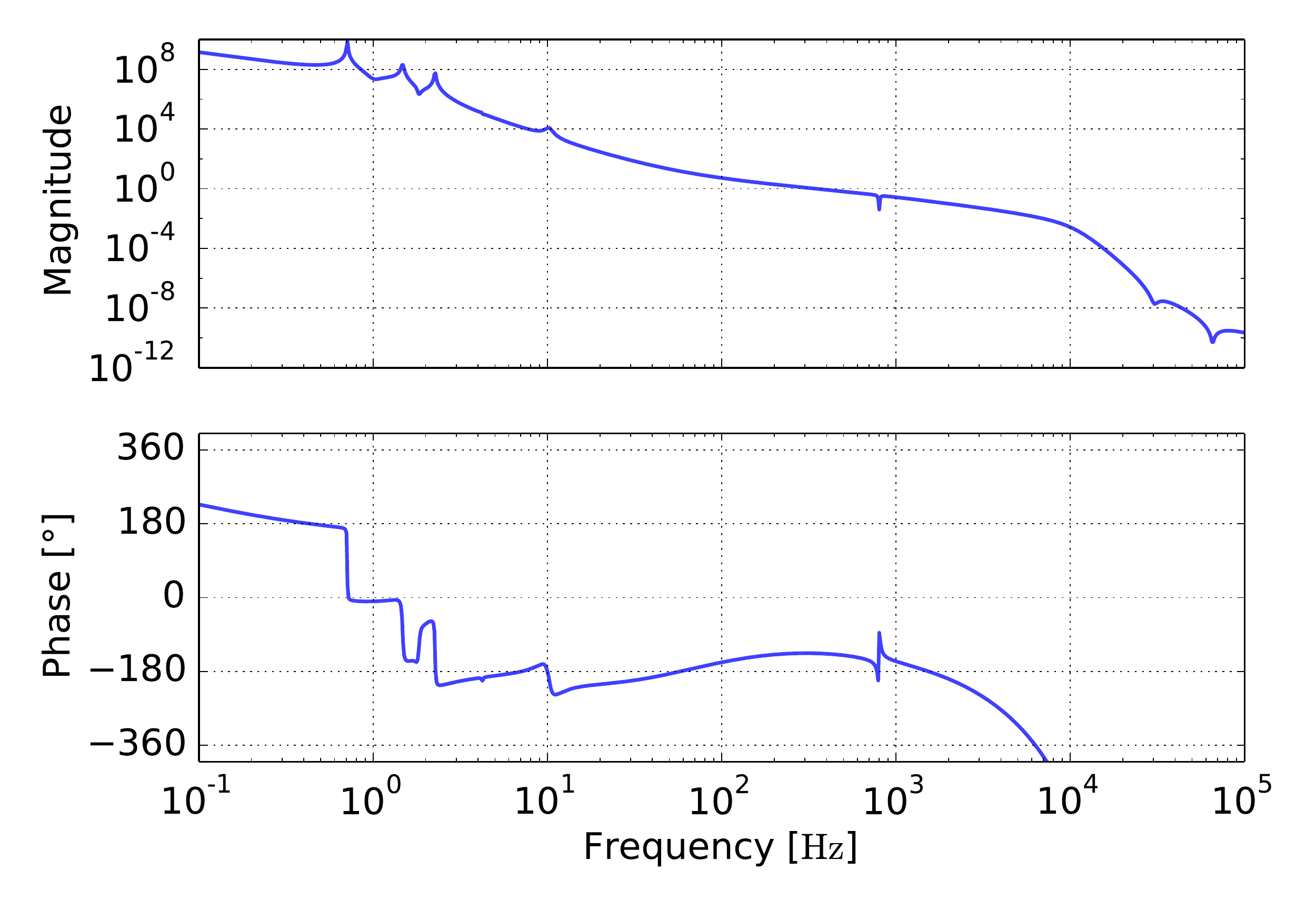}
    \caption{\label{fig:open-loop-gain}Simulated \ssm{} controller open loop gain. The majority of the gain is applied to correct displacements due to seismic noise below \SI{10}{\hertz}. The unity gain frequency is \SI{350}{\hertz} and the phase margin is \SI{44}{\degree}.}
  \end{figure}

  The open loop gain of the control system is shown in Figure\,\ref{fig:open-loop-gain}. The greatest gain is required in the region below \SI{10}{\hertz} where the test mass suspensions provide little to no isolation from seismic noise. Above this frequency the required gain decreases as other noise sources become dominant. The unity gain frequency is at \SI{350}{\hertz}.
  
\section{\label{app:parameters}Experimental parameters}

  The parameters used in the simulations presented in this work are shown in Table\,\ref{tab:parameters}. Unless otherwise stated, the mirrors specified in the figures and simulations are assumed to have unity reflectivity. All listed transmissivities represent power, no substrate loss is assumed for any optic and all simulations have been performed using the plane-wave approximation.
  
  \begin{table}
    \begin{ruledtabular}
      \begin{tabular}{ll}
	\textbf{Parameter}   & \textbf{Fiducial value} \\
	Laser wavelength $\lambda_{0}$        & \SI{1064}{\nano\meter} \\
	Input power             & \SI{1.8}{\watt} \\
	Round-trip cavity length $L_{\textrm{RT}}$ & \SI{2.83}{\meter} \\
	\mi{} transmissivity & \SI{e4}{} ppm \\
	ITM transmissivity      & \SI{700}{} ppm                 \\
	Arm cavity full-width half-maximum & \SI{12.2}{\kilo\hertz} \\
	Arm cavity finesse      & \SI{8663}{} \\
	BHD quantum efficiency  & \SI{0.95}{\ampere\per\watt} \\
	PDH quantum efficiency  & \SI{0.80}{\ampere\per\watt} \\
	\rt{}                   & \SI{10}{\kilo\ohm} \\
	PDH demodulation gain   & \SI{21}{\decibel} \\
	ADC/DAC quantisation noise  & \SI{1.8}{\micro\volt\per\sqrthz} \\
	ETM mass                & \SI{113}{\gram} \\
	ETM fibres              & \SI{4}{} \\
	ETM fibre diameter      & \SI{40}{\micro\meter} \\
	ETM fibre length        & \SI{200}{\milli\meter} \\
	ITM mass                & \SI{0.86}{\gram} \\
	ITM fibres              & \SI{2}{} \\
	ITM fibre diameter      & \SI{10}{\micro\meter} \\
	ITM fibre length        & \SI{100}{\milli\meter} \\
	Suspension vertical-to-horizontal coupling & \SI{0.01}{} \\
      \end{tabular}
      \caption{\label{tab:parameters}Experimental parameters. The properties for the suspensions and test masses are given in order for the reader to be able to reproduce the suspension thermal noise spectral density presented in Figure\,\ref{fig:noise-budget}.}
    \end{ruledtabular}
    \end{table}

\bibliography{control}

\end{document}